\documentclass[conference]{IEEEtran}
\IEEEoverridecommandlockouts

\usepackage[letterpaper, left=1in, right=1in, bottom=1in, top=0.75in]{geometry}

\usepackage{algorithmic}
\usepackage{amsthm}
\usepackage{amsmath}
\usepackage{amssymb}
\usepackage{amsfonts}
\usepackage{balance}
\usepackage{caption}
\usepackage{cite}
\usepackage{color}
\usepackage{esint}
\usepackage{graphicx}
\usepackage{epstopdf}
\usepackage{subcaption}

\usepackage{epstopdf}

\usepackage{textcomp}
\def\BibTeX{{\rm B\kern-.05em{\sc i\kern-.025em b}\kern-.08em
    T\kern-.1667em\lower.7ex\hbox{E}\kern-.125emX}}
    
\theoremstyle{plain}
\newtheorem{thm}{\protect\theoremname}

\makeatother

\providecommand{\theoremname}{Theorem}

\theoremstyle{plain}
\newtheorem{lem}{\protect\lemmaname}

\makeatother

\providecommand{\lemmaname}{Lemma}

\theoremstyle{plain}

\providecommand{\propositionname}{Proposition}

\makeatother

\theoremstyle{plain}
\newtheorem{cor}{\protect\corname}

\makeatother

\providecommand{\corname}{Corollary}

\DeclareMathOperator{\Csc}{csc}    
    
\begin{document}

\title{Ergodic capacity analysis of reconfigurable intelligent surface assisted wireless systems
\thanks{This work has received funding from the European Commission’s Horizon 2020 research and innovation programme under grant agreement No. 761794.}
}

\author{\IEEEauthorblockN{ Alexandros--Apostolos A. Boulogeorgos,  and Angeliki Alexiou}
\IEEEauthorblockA{\textit{Department of Digital Systems,
University of Piraeus}, 
Piraeus, Greece \\
E-mails: al.boulogeorgos@ieee.org, alexiou@unipi.gr}
}

\maketitle

\begin{abstract}  
	This paper presents the analytic framework for evaluating the ergodic capacity (EC) of  the reconfigurable intelligent surface (RIS) assisted systems. 
	Moreover, high-signal-to-noise-ratio  and high-number of reflection units (RUs) approximations for the  EC are provided.   
	Finally,  the special case in which the RIS is equipped with a single RU is investigated. 
	Our analysis is verified through respective Monte Carlo simulations, which highlight the accuracy of the proposed framework. 
\end{abstract}



\maketitle

\section{Introduction}\label{S:Intro}

While the wireless world moves towards the sixth generation (6G) era, the
 data-rate and network traffic demands have been exponential increased~\cite{A:A_vision_of_6G_wireless_systems,A:LC_CR_vs_SS,ref1_VTC,WP:Wireless_Thz_system_architecture_for_networks_beyond_5G}. 
Technological advances, 
such as massive multiple-input multiple-output, full-duplexing, and high-frequency communications, have been advocated, due to the power consumption increase that they cause\cite{A:IRS_for_multicell_MIMO_communications,A:ED_in_FD_with_Residual_RF_impairments,C:ADistanceAndBWDependentAdaptiveModulationSchemeForTHzCommunications,C:UserAssociationInUltraDenseTHzNetworks}, as well as their performance limitations when operating in unfavorable wireless propagation environment~\cite{PhD:Boulogeorgos,A:LIS_assisted_wireless_communication_exploiting_statistical_CSI,C:Analytical_performance_evaluation_of_THz_Wireless_Fiber_Externders,PIMRC_275-400,Boulogeorgos2020}.    

To surpass the aforementioned issues, the use of reconfigurable intelligent surfaces (RISs) in order to exploit  the implicit randomness of the propagation environment have attacted the attention of both academia and industry~\cite{A:Smart_radio_enviroments}. Most RISs consist of two dimensional reflection units (RUs) arrays, which are controlled by at least one micro-controller, and can alter the incoming electromagnetic (EM) field~\cite{A:Exploration_of_intercell_wireless_millimeter_wave_communication_in_the_landscape_of_intelligent_metasurfaces}. In particular, each RU can independently change the phase of the incident EM wave; hence,  they are able to collaboratively create a favorable wireless channel~\cite{C:Intelligent_reflecting_surface_enhanced_wireless_network}.

Scanning the open literature, the perfoamance analysis of RIS-assisted systems is a topic of much hype~(see e.g.,\cite{	A:Wireless_communications_through_RIS,C:Transmission_throug_LIS,Thirumavalavan2020,Jung2019,A:RIS_vs_Relaying} and references therein). 
Specifically, in~\cite{A:Wireless_communications_through_RIS} and~\cite{C:Transmission_throug_LIS,Bjornson2020}, the authors provided a symbol error rate (SER) bound for RIS-assisted systems. 
Note that these upper-bounds are quite tight for RIS utilizations with high-number of RUs, but, in the low-RUs regime, they are not so~accurate.
Similarly, in~\cite{Thirumavalavan2020}, an error analysis was provided for RIS-assisted non-orthogonal multiple access systems. Again, the authors employed the central limit theorem for approximating the distribution of the equivalent wireless channel. 
As a consequence, the results are accurate only for scenarios, in which the RIS consists of a large-number of RUs. 
In~\cite{Jung2019}, the authors presented an asymptotic analysis of the uplink sum-rate of a RIS-assisted system, assuming that the established channels follow Rician distribution.  
Finally, in~\cite{Bjornson2020},  the performance of such systems in terms of energy efficiency was quantified.

To the best of the authors knowledge, no analytical assessment of the EC in RIS-assisted systems  has been reported. 
Motivated by this, this work presents the analytical framework that quantifies the EC of RIS-assisted systems. 
In this sense, we initially present novel a closed-form expression for the probability density function (PDF)  of the end-to-end (e2e)  fading channel coefficient of the RIS-assisted  system. Moreover, the PDF of the e2e  fading channel for the special case, in which the RIS is equipped with a single RU, is also presented. Building upon them, we extract closed-form expressions for the EC of the RIS-assisted system for both cases, in which the RIS is equipped with multiple and a single RU. Finally, tight novel high-SNR and high-RU number approximations for the EC are derived.

\subsubsection*{Notations}
The operators $\mathbb{E}[\cdot]$, $\mathbb{V}[\cdot]$ and $|\cdot|$ respectively denote the statistical expectation, variance, and the absolute value, whereas $\exp\left(x\right)$ and $\log_2\left(x\right)$ respectively stand for the exponential and the binary logarithmic functions.
Additionally, $\ln\left(x\right)$ refers to the natural logarithm of $x$, while $\sqrt{x}$ and $\displaystyle{\lim_{x\to a}}\left(f(x)\right)$ respectively return the square root of $x$ and the limit of the function $f(x)$ as $x$ tends to $a$.  
Furthermore,  $(x)_n$  denotes the Pochhammer operator. 
Also, $\csc(x)$ and $\mathrm{sec}(x)$ respectively return the cosecant and the secant of $x$. The upper and lower incomplete Gamma functions~\cite[eq. (8.350/2), (8.350/3)]{B:Gra_Ryz_Book} are respectively denoted by $\Gamma\left(\cdot, \cdot\right)$ and $\gamma\left(\cdot, \cdot\right)$, while the Gamma function is represented by $\Gamma\left(\cdot\right)$~\cite[eq. (8.310)]{B:Gra_Ryz_Book}, whereas $K_v(\cdot)$ and $I_v(\cdot)$ are respectively the modified Bessel function of the second~\cite[eq. (9.6.2)]{B:Abramowitz} and first kind of order $v$~\cite[eq. (9.6.3)]{B:Abramowitz}.  
Moreover, $F_0(\cdot)$, $\mathrm{E}(\cdot)$, and $\mathrm{K}(\cdot)$ respectively represent the polygamma function of the zero order~\cite[eq. (6.4.1)]{B:Abramowitz}, the elliptic integral function~\cite[eq. (17.1.1)]{B:Abramowitz}, and the complete elliptic integral function of the first kind~\cite[eq. (17.3.1)]{B:Abramowitz}.  
Furthermore, $\,_2F_1(\cdot,\cdot;\cdot; \cdot)$ stands for the Gauss hypergeometric function~\cite[eq. (4.1.1)]{B:Abramowitz}, while $\,_pF_q\left(a_1,\cdots,a_p; b_1, \cdots, b_q; x\right)$ is the generalized hypergeometric function~\cite[eq. (9.14/1)]{B:Gra_Ryz_Book}. 
Meanwhile, $U\left(a, b, x\right)$ and $G_{p, q}^{m, n}\left(x\left| \begin{array}{c} a_1, a_2, \cdots, a_{p} \\ b_{1}, b_2, \cdots, b_q\end{array}\right.\right)$ respectively represent the confluent hypergeometric function of second kind~\cite[ch. 9.2]{B:Gra_Ryz_Book}, and the Meijer's G-function~\cite[eq. (9.301)]{B:Gra_Ryz_Book}.

\section{System  Model}\label{sec:SM}

\begin{figure}
	\centering\includegraphics[width=1\linewidth,trim=0 0 0 0,clip=false]{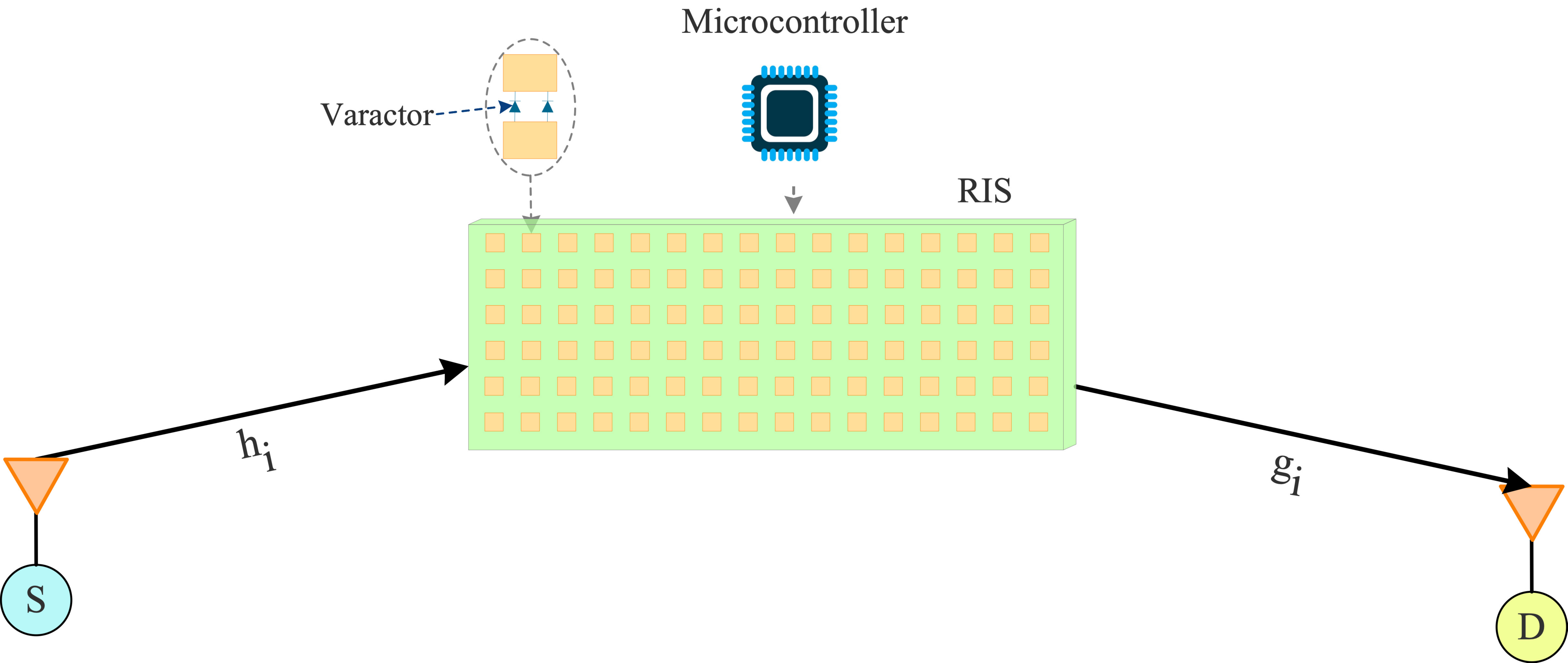}
	\caption{System model of the RIS-assisted wireless system.}
	\label{fig:SM}
\end{figure}

As shown in Fig.~\ref{fig:SM}, we consider a RIS-assisted wireless system, in which  a single-antenna source (S) node communicates with a single-antenna destination (D) node through a RIS, that consists of $N$ RUs. The baseband equivalent channels between S and the $i$-th RU of the RIS, $h_i$, as well as the one between the $i-$th RU and D, $g_i$, are assumed to be independent and Rayleigh distributed random variables (RVs) with scale parameters $1$. {This assumption originates from the fact that even if the line-of-sight links between S-RIS and RIS-D are blocked, there still exist extensive~scatters. Moreover, as usual practice, in this work, we also neglect the deterministic path-gain in the fading coefficients $h_i$ and $g_i$. } 

Hence, the baseband equivalent received signal at D is given~by
\begin{align}
y = \sum_{i=1}^{N} h_i g_i r_i x + n,
\label{Eq:y}
\end{align}    
where $n$ denotes the additive white Gaussian noise and can be modeled as a zero-mean complex Gaussian RV with variance equal $N_o$. Likewise, $r_i$ represents the $i-$th RU response and can be expressed~as
\begin{align}
r_i = |r_i| \exp\left(j\theta_i\right),
\label{Eq:ri}
\end{align}
with $\theta_i$ being the PS applied by the $i-$th reflecting RU of the RIS. In this work, we assume that the phases of the channels $h_i$ and $g_i$ are perfectly known to the RIS, and that thus the RIS selects the optimal phase shifting, which is 
$\theta_i = -\left(\phi_{h_i} + \phi_{g_i}\right)$,
where $\phi_{h_i}$ and $ \phi_{g_i}$ are respectively the phases of  $h_i$ and $g_i$.  
In addition, without loss of generality, it is assumed that the reflected gain of the $i-$th RU, $|r_i|$, is equal to $1$. Hence,~\eqref{Eq:ri} can be simplified~as
\begin{align}
r_i = \exp\left(-j\left(\phi_{h_i} + \phi_{g_i}\right)\right).
\label{Eq:ri_s1}
\end{align}
By employing~\eqref{Eq:ri_s1},~\eqref{Eq:y} can be expressed~as
\begin{align}
y = A x + n,
\label{Eq:y_s1}
\end{align}
where 
\begin{align}
A = \sum_{i=1}^{N} |h_i| |g_i|. 
\label{Eq:A}
\end{align}

\section{Performance Analysis}\label{sec:PA}


The following theorem returns closed-form approximation for the PDF and CDF of $A$. 
\begin{lem}
	The PDF of $A$ can be  evaluated~as 
	\begin{align}
	f_{A}(x) = \frac{x^a}{b^{a+1}\Gamma(a+1)} \exp\left(-\frac{x}{b}\right),
	\label{Eq:f_A}
	\end{align}
	where
	\begin{align}
	a = \frac{k_1^2}{k_2}-1,
	\text{ and }
	b = \frac{k_2}{k_1},
	\label{Eq:b}
	\end{align}
	with 
	\begin{align}
	k_1 = \frac{N\pi}{2},
	\text{ and }
	k_2 = 4N\left(1-\frac{\pi^2}{16}\right).
	\label{Eq:k2}
	\end{align}
\end{lem}
\begin{IEEEproof}
	Please refer to Appendix A. 
\end{IEEEproof}

\underline{Special case}: For the case in which the RIS consists of a single RU, i.e. $N=1$, $A$ is the product of two independent and identical Rayleigh distributed random variables (RVs); thus, it follows a double Rayleigh distribution and its PDF  can  be  obtained as~\cite[eq. (3)]{C:Investigations_of_outdoor_to_indoor_mobile_to_mobile_radio_communication_channelss}
\begin{align}
f_{A}^{s}= x K_0\left(x\right).
\label{Eq:f_A_special_case}
\end{align}


The following theorem return a novel closed-form expression for the EC. 
\begin{thm}
	The EC of the RIS-assisted system can be analytically computed as in~\eqref{Eq:C6}, given at the top of the next page. 
	\begin{figure*}
		\begin{align}
		C&= 
		\frac{a^2- a }{(a-1)_2}  \log_2\left({b^2}{\rho_s}\right)
		+ \frac{2 \left(a^2-a\right)}{\ln(2) (a-1)_2} F_0\left(3+a\right)
		+ \frac{\pi \Csc\left(\frac{a\pi}{2}\right) \,_1F_2\left(1+\frac{a}{2}; \frac{3}{2}, 2+ \frac{a}{2}, -\frac{1}{4 b^2 \rho_s} \right)}{\ln(2) (2+a) b^{a+2} \Gamma(a+1) \rho_s^{\frac{a}{2}+1}} 
		\nonumber \\ &
		+ \frac{\pi \mathrm{sec}\left(\frac{a \pi}{2}\right) \,_1F_2\left(\frac{a+1}{2};\frac{1}{2},\frac{a+3}{2}, - \frac{1}{4 b^2 \rho_s}\right)}{(a+1) b^{a+1}\ln\left(2\right)\Gamma(a+1) \rho_s^{\frac{a+1}{2}}} 
		+ \frac{\,_2F_3\left(1,1;2,1-\frac{a}{2},\frac{3-a}{2}, - \frac{1}{4 b^2 \rho_s}\right)}{\ln(2)(a-1)_2 b^2 \rho_s}   
		\label{Eq:C6}
		\end{align}
		\hrulefill
	\end{figure*} 
In~\eqref{Eq:C6}, $\rho_t=\frac{P_t}{N_o}$, where $P_t$ is the S transmission power. 
\end{thm}
\begin{IEEEproof}
	Please refer to Appendix B. 
\end{IEEEproof}

The following corollaries present high-SNR and high-$N$ approximations for the EC. 
\begin{cor}
	In the high SNR regime, the EC can be approximated as in~\eqref{Eq:Capacity_approximation}, given at the top of the next~page.
	\begin{figure*}
		\begin{align}
		C_{\rho}&\approx 
		\frac{1}{\ln(2)(a-1)_2 b^2 \rho_t}  
		+ \frac{\left(a^2- a\right) \log_2\left({b^2}{\rho_t}\right)
			+ \frac{2 \left(a^2-a\right)}{\ln(2) (a-1)_2} F_0\left(3+a\right) }{(a-1)_2}  
		\nonumber \\ &
		+ \frac{\pi \Csc\left(\frac{a\pi}{2}\right) }{\ln(2) (2+a) b^{a+2} \Gamma(a+1) \rho_t^{\frac{a}{2}+1}} 
		+ \frac{\pi \mathrm{sec}\left(\frac{a \pi}{2}\right)}{(a+1) b^{a+1}\ln\left(2\right)\Gamma(a+1) \rho_t^{\frac{a+1}{2}}} 
		\label{Eq:Capacity_approximation}
		\end{align}
		\hrulefill
	\end{figure*} 
\end{cor}
\begin{IEEEproof}
	For $\rho_t\to \infty$,
	$y =  \frac{1}{4 b^2 \rho_t} \to 0$. Moreover,
	\begin{align}
	\lim_{y\to 0}& \,_1F_2\left(1+\frac{a}{2}; \frac{3}{2}, 2+ \frac{a}{2}, -y \right) = 1,\\
	\lim_{y\to 0} &  \,_1F_2\left(\frac{a+1}{2};\frac{1}{2},\frac{a+3}{2}, - y \right) = 1
	\end{align} 
	and
	\begin{align}
	\lim_{y\to 0}  \,_2F_3\left(1,1;2,1-\frac{a}{2},\frac{3-a}{2}, - y\right) =1.
	\end{align}
	Thus,in the high SNR regime~\eqref{Eq:C6} can be approximated as in~\eqref{Eq:Capacity_approximation}. This concludes the proof. 
\end{IEEEproof}

\begin{cor}
	In the high SNR and $N$ regime, the EC can be approximated~as
	\begin{align}
	C_{\rho,N}&\approx 
	\frac{1}{\ln(2)(a-1)_2 b^2 \rho_t}  
	+ \frac{a^2- a }{(a-1)_2}  \log_2\left({b^2}{\rho_t}\right)
	\nonumber \\ & 
	+ \frac{2 \left(a^2-a\right)}{\ln(2) (a-1)_2} F_0\left(3+a\right).
	\label{Eq:Capacity_approximation2}
	\end{align}
\end{cor}
\begin{IEEEproof}
	In the high SNR regime, as $N\to\infty$, $a\to\infty$; hence, since $\Gamma\left(a+1\right)$ is an increasing function, as $N\to\infty$, $\Gamma\left(a+1\right)\to\infty$, or equivalently $\frac{1}{\Gamma\left(a+1\right)}\to 0$. This indicates 
	$\lim_{N\to \infty} \mathcal{B}_1 = \lim_{N\to \infty} \mathcal{B}_2 = 0$, where 
	\begin{align}
	\mathcal{B}_1 = \frac{\pi}{\ln(2) (2+a) b^{a+2} \Gamma(a+1) \rho_t^{\frac{a}{2}+1}} \Csc\left(\frac{a\pi}{2}\right)
	\end{align}
	and
	\begin{align}
	\mathcal{B}_2 = \frac{\pi}{(a+1) b^{a+1}\ln\left(2\right)\Gamma(a+1) \rho_t^{\frac{a+1}{2}}} \mathrm{sec}\left(\frac{a \pi}{2}\right).
	\end{align}
	 Therefore,~\eqref{Eq:Capacity_approximation} can be approximated as in~\eqref{Eq:Capacity_approximation2}. This concludes the proof. 
\end{IEEEproof}

\underline{Special case}: For $N=1$, the EC can be evaluated according to the following lemma. 
\begin{lem}
	For a single RU RIS, the EC can be obtained~as
	\begin{align}
	C_s &= \frac{1}{8\ln(2)\rho_t^2} G_{1,3}^{3,1}\left(\frac{1}{4\rho_t^2}\left|\begin{array}{c} -1 \\ -1, -1, 0\end{array}\right. \right) 
	\nonumber \\ &
	- \frac{1}{4\ln(2) \rho_t} G_{1,3}^{3,1}\left(\frac{1}{4\rho_t^2}\left|\begin{array}{c} -\frac{1}{2} \\ -\frac{1}{2}, -\frac{1}{2}, -\frac{1}{2}\end{array}\right. \right)
	\nonumber \\ &
	+ \frac{1}{8\ln(2)\rho_t^2} G_{2,4}^{4,1}\left(\frac{1}{4\rho_t^2}\left|\begin{array}{c} -1, 0 \\ -1, -1, -1, 1\end{array}\right.\right).
	\label{Eq:Cs_final}
	\end{align}
\end{lem}
\begin{IEEEproof}
	Please refer to Appendix C. 
\end{IEEEproof}

\section{Numerical Results}\label{sec:Results}

This section is focused on  verifying the theoretical framework through respective Monte Carlo simulations and reporting the EC performance of the RIS-assisted  system. Unless otherwise stated, in what follows, we use continuous lines and markers to respectively denote theoretical and simulation~results. 

\begin{figure}
	\centering\includegraphics[width=0.8\linewidth,trim=0 0 0 0,clip=false]{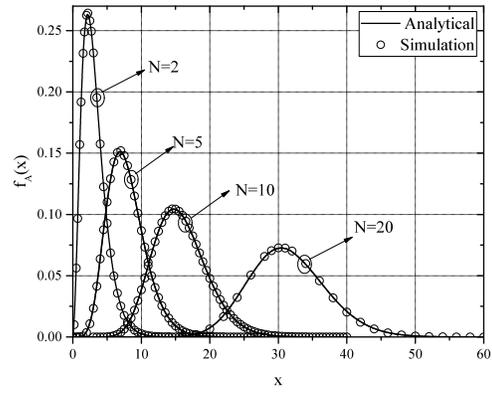}
	\caption{The PDF of the equivalent e2e channel for different $N$.}
	\label{fig:PDF_A}
\end{figure}

Figure~\ref{fig:PDF_A} illustrates the PDF of the equivalent e2e channel of the RIS-assisted  system, for different number of RUs. From this figure, it is observed that the theoretical and simulation results coincide; thus, verifying the presented analytical framework. Additionally, it is observed that, as $N$ increases, the equivalent e2e channel values also increase. This indicates that by increasing $N$, the diversity gain of the RIS-assisted~system improves.

\begin{figure}
	\centering\includegraphics[width=0.8\linewidth,trim=0 0 0 0,clip=false]{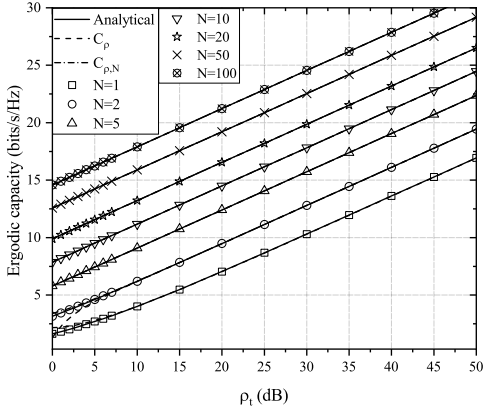}
	\caption{Capacity vs ${\rho_{t}}$, for different values of~$N$.}
	\label{fig:Capacity}
\end{figure} 

Figure~\ref{fig:Capacity} depicts the EC as a function of $\rho_t$, for different values of $N$. In this figure, continuous lines denote the analytical results, the dashed ones stand for the high-SNR approximation, while the dashed-dotted ones represent the high SNR-and-$N$ approximation.  We observe that  both the high-SNR and the high-SNR-$N$ approximations provides excellent fits even in the medium and low transmission SNR regimes.  Likewise, it is obvious that, for a fixed $N$, as $\rho_t$ increases, the EC also increases. For instance, for $N=2$, as $\rho_t$ changes from $5$ to $10\text{ }\mathrm{dB}$, the EC improves for about $34.2\%$. Moreover, for a given $\rho_t$, as $N$ increases, the achievable EC increases. For example, for $\rho_t = 10\text{ }\mathrm{dB}$, as $N$ shifts from $50$ to $100$, the EC increases for approximately $12.64\%$. Finally, this figure reveals that, independently of $\rho_t$, as $N$ doubles, the EC increases for about $2\text{ }\mathrm{bits/s/Hz}$. 

\section{Conclusions}\label{sec:Conclusions}

This contribution studied the EC of RIS-assisted systems. 
 After obtaining the PDF of the equivalent e2e  channel of the RIS-assisted system, we  extracted novel closed-form expressions for the EC together with low-complexity tight high-SNR and high-$N$ approximations.      
The analytical results were compared against respective Monte Carlo simulations, which validated their accuracy and revealed that as the number of RUs increases, the EC also increases.    
Interestingly, it was observed that as the number of RUs doubles, the EC for about $2\text{ }\mathrm{bits/s/Hz}$.

\section*{Appendices}

\section*{Appendix A}
\section*{Proof of Lemma 1}

Since,  $|h_i|$ and $|g_i|$ are Rayleigh distributed RVs, from~\eqref{Eq:A}, it becomes evident that $A$ is the sum of $N$ independent and identical double Rayleigh processes, which PDF, according to~\cite[ch. 2.2.2]{B:Stochastic_methods}, can be tightly approximated as the first term of a Laguerre series expansion, i.e.,~\eqref{Eq:f_A}. The parameters $a$ and $b$ are given in~\eqref{Eq:b}, whereas $k_1$ and $k_2$ can be evaluated as~\cite[eq. (2.74)]{B:Stochastic_methods}
\begin{align}
k_1 = \mathbb{E}[A],
\label{Eq:k_1_def}
\end{align}
and
\begin{align}
k_2 = 4 \mathbb{V}[A].
\label{Eq:k_2_def}
\end{align}

The expected value of $A$ can be obtained~as
\begin{align}
\mathbb{E}[A] = \sum_{i=1}^{N} \mathbb{E}\left[|h_i| |g_i|\right],
\label{Eq:EA_1}
\end{align}
or, due the $|h_i|$ and $|g_i|$ independency,
\begin{align}
\mathbb{E}[A] = \sum_{i=1}^{N} \mathbb{E}\left[|h_i|\right] \mathbb{E}\left[|g_i|\right].
\label{Eq:EA_2}
\end{align} 
Likewise, $|h_i|$ and $|g_i|$ follow Rayleigh distribution with variances $1$; thus,
\begin{align}
\mathbb{E}\left[|h_i|\right] = \mathbb{E}\left[|g_i|\right] = \sqrt{\frac{\pi}{2}}. 
\label{Eq:Ehi}
\end{align}
By substituting~\eqref{Eq:Ehi} into~\eqref{Eq:EA_2}, we get
\begin{align}
\mathbb{E}[A] = N\frac{\pi}{2}.
\label{Eq:EA_3}
\end{align}
 
Similarly, the variance of $A$ can be computed~as
\begin{align}
\mathbb{V}[A] = N\left(1-\frac{\pi^2}{16}\right).
\label{Eq:VA_1} 
\end{align}
By substituting~\eqref{Eq:EA_3} and~\eqref{Eq:VA_1} into~\eqref{Eq:k_1_def} and~\eqref{Eq:k_2_def}, we obtain~\eqref{Eq:k2}. This concludes the~proof.

\section*{Appendix B}
\section*{Proof of Theorem 1}

The EC is defined~as
\begin{align}
C = \mathbb{E}\left[ \log_2\left(1+\rho\right) \right],
\label{Eq:C1}
\end{align}
which can be equivalently written as
\begin{align}
C = \int_{0}^{\infty} \log_2\left(1+\rho_t y^2\right) f_{A}(y)\text{ }\mathrm{dy}.
\label{Eq:C2}
\end{align}
By substituting~\eqref{Eq:f_A} into~\eqref{Eq:C2}, the EC can be rewritten~as
\begin{align}
C= \frac{x^a}{b^{a+1}\Gamma(a+1)} \int_{0}^{\infty} \exp\left(-\frac{y}{b}\right) \log_2\left(1+\rho_t y^2\right) \text{ }\mathrm{dy},
\end{align}
or 
\begin{align}
C= \frac{1}{b^{a+1}\ln\left(2\right)\Gamma(a+1)} \mathcal{K},
\label{Eq:C4}
\end{align}
where
\begin{align}
\mathcal{K} = \int_{0}^{\infty} y^a \exp\left(-\frac{y}{b}\right) \ln\left(1+\rho_t y^2\right) \text{ }\mathrm{dy}.
\label{Eq:K}
\end{align}
Based on~\cite[eq. (15.1.1)]{B:Abramowitz},~\eqref{Eq:K} can be written~as
\begin{align}
\mathcal{K} = \rho_t \int_{0}^{\infty} y^{a+2} \exp\left(-\frac{y}{b}\right) \,_2F_1\left(1, 1;2; -\rho_t y^2\right) \text{ }\mathrm{dy}.
\label{Eq:K2},
\end{align}
which, after applying integration by parts as well as~\cite[eq. (07.23.21.0015.01)]{Hypergeometric}, can be expressed as in~\eqref{Eq:K3}, given at the top of the following page.
\begin{figure*}
	\begin{align}
	&\mathcal{K} = 
	4 a b^{a+3} \Gamma(a) \rho_t \ln\left({b}{\sqrt{\rho_t}}\right)
	+ 6 a^2 b^{a+3} \Gamma(a) \rho_t \ln\left({b}{\sqrt{\rho_t}}\right)
	+ 2 a^3 b^{a+3} \Gamma(a) \rho_t \ln\left({b}{\sqrt{\rho_t}}\right)
	\nonumber \\ &
	+ 2 a (a+1) (a+2) b^{a+3} \Gamma(a) F_0\left(3+a\right)
	- \frac{\pi}{(4+a) b \rho_t^{\frac{a}{2}+1}} \Csc\left(\frac{a\pi}{2}\right) \,_1F_2\left(2+\frac{a}{2}; \frac{3}{2}, 3+ \frac{a}{2}, -\frac{1}{4 b^2 \rho_t} \right)
	\nonumber \\ &
	- \frac{\pi}{(a+3) \rho_t^{\frac{a+1}{2}}} \mathrm{sec}\left(\frac{a \pi}{2}\right) \,_1F_2\left(\frac{a+3}{2};\frac{1}{2},\frac{a+5}{2}, - \frac{1}{4 b^2 \rho_t}\right)
	+ a b^{a+1} \Gamma(a) \,_2F_3\left(1,1;2,\frac{1-a}{2},-\frac{a}{2}, - \frac{1}{4 b^2 \rho_t}\right)
	\label{Eq:K3} 
	\end{align}
	\hrulefill
\end{figure*}
In addition, by substituting~\eqref{Eq:K3} into~\eqref{Eq:C4}, and after some algebraic manipulations, we extract~\eqref{Eq:C5}, given at the top of the following page. 
\begin{figure*}
	\begin{align}
	C&= 
	4 a b^{2} \frac{\Gamma(a)}{\Gamma(a+1)} \rho_t \log_2\left({b}{\sqrt{\rho_t}}\right)
	+ 6 a^2 b^{2} \frac{\Gamma(a)}{\Gamma(a+1)}\rho_t \log_2\left({b}{\sqrt{\rho_t}}\right)
	+ 2 a^3 b^{2} \frac{\Gamma(a)}{\Gamma(a+1)} \rho_t \log_2\left({b}{\sqrt{\rho_t}}\right)
	\nonumber \\ &
	+ \frac{2}{\ln(2)} a (a+1) (a+2) b^{2} \frac{\Gamma(a)}{\Gamma(a+1)} F_0\left(3+a\right)
	\nonumber \\ &
	- \frac{\pi}{\ln(2) (4+a) b^{a+2} \Gamma(a+1) \rho_t^{\frac{a}{2}+1}} \Csc\left(\frac{a\pi}{2}\right) \,_1F_2\left(2+\frac{a}{2}; \frac{3}{2}, 3+ \frac{a}{2}, -\frac{1}{4 b^2 \rho_t} \right)
	\nonumber \\ &
	- \frac{\pi}{(a+3) b^{a+1}\ln\left(2\right)\Gamma(a+1) \rho_t^{\frac{a+1}{2}}} \mathrm{sec}\left(\frac{a \pi}{2}\right) \,_1F_2\left(\frac{a+3}{2};\frac{1}{2},\frac{a+5}{2}, - \frac{1}{4 b^2 \rho_t}\right)
	\nonumber \\ &
	+ \frac{a}{\ln(2)}  \frac{\Gamma(a)}{\Gamma(a+1)} \,_2F_3\left(1,1;2,\frac{1-a}{2},-\frac{a}{2}, - \frac{1}{4 b^2 \rho_t}\right)
	\label{Eq:C5}
	\end{align}
	\hrulefill
\end{figure*}
Finally, by taking into account that
$\frac{\Gamma(x+n)} {\Gamma(x)} = (x)_n$,~ 
\eqref{Eq:C5} can be rewritten as in~\eqref{Eq:C6}. 
This concludes the~proof.

\section*{Appendix C}
\section*{Proof of Lemma 2}

Based on~\eqref{Eq:C1}, the EC can be obtained~as
\begin{align}
C_s = \int_{0}^{\infty} \log_2\left(1+\rho_t x\right) f_{\rho}^{s}(x)\text{ }\mathrm{dx},
\end{align}
or equivalently
\begin{align}
C_s = \frac{1}{\ln(2)} \int_{0}^{\infty} \ln\left(1+\rho_t x^2\right) f_{A}^{s}(x)\text{ }\mathrm{dx},
\end{align}
which, with the aid of~\eqref{Eq:f_A}, can be expressed~as
\begin{align}
C_s = \frac{1}{4\ln(2)\rho_t} \mathcal{C}_1 - \frac{1}{2\ln(2)\sqrt{\rho_t}} \mathcal{C}_2 + \frac{1}{4\ln(2)\rho_t} \mathcal{C}_3,
\label{Eq:Cs_step1}
\end{align}
where 
\begin{align}
\mathcal{C}_1 &= \int_{0}^{\infty} K_0\left(\sqrt{ \frac{x}{\rho_t}}\right) \ln\left(1 + \rho_t x\right) \text{ }\mathrm{dx},\label{Eq:C_1} \\
\mathcal{C}_2 &= \int_{0}^{\infty} x^{-1/2} K_1\left(\sqrt{ \frac{x}{\rho_t}}\right) \ln\left(1 + \rho_t x\right) \text{ }\mathrm{dx}\label{Eq:C_2}
\end{align}
and
\begin{align}
\mathcal{C}_3 &= \int_{0}^{\infty} K_2\left(\sqrt{ \frac{x}{\rho_t}}\right) \ln\left(1 + \rho_t x\right) \text{ }\mathrm{dx}.
\label{Eq:C_3}
\end{align}
Moreover, by using~\cite[eq. (8.352/2)]{B:Gra_Ryz_Book},~\eqref{Eq:C_1}-\eqref{Eq:C_3} can be equivalently written~as
\begin{align}
\mathcal{C}_1 &= \rho_t \int_{0}^{\infty} x K_0\left(\sqrt{ \frac{x}{\rho_t}}\right) \,_2F_1\left(1,1;2;\rho_t x\right) \text{ }\mathrm{dx},\label{Eq:C_1_step2} \\
\mathcal{C}_2 &= \rho_t \int_{0}^{\infty} x^{1/2} K_1\left(\sqrt{ \frac{x}{\rho_t}}\right) \,_2F_1\left(1,1;2;\rho_t x\right) \text{ }\mathrm{dx}\label{Eq:C_2_step2}
\end{align}
and
\begin{align}
\mathcal{C}_3 &= \rho_t \int_{0}^{\infty} x K_2\left(\sqrt{ \frac{x}{\rho_t}}\right) \,_2F_1\left(1,1;2;\rho_t x\right) \text{ }\mathrm{dx}.
\label{Eq:C_3_step2}
\end{align}
Moreover, with the aid of~\cite[eq. (03.04.26.0009.01)]{S:Bessel_K_to_Meijer} and~\cite[eq. (17)]{A:The_algorithm_for_calculating_integrals_of_hypergeometric_type_functions_and_its_relaization_in_REDUCE_system},~\eqref{Eq:C_1_step2}-\eqref{Eq:C_3_step2} can be respectively written~as
\begin{align}
\mathcal{C}_1 &= \frac{\rho_t}{2} \int_{0}^{\infty} x  G_{0,2}^{2,0}\left(\left.\frac{x}{4\rho_t}\right| 0, 0 \right) 
\nonumber \\ & \hspace{+2.7cm}\times
G_{2,2}^{1,2}\left(\rho_t x\left|\begin{array}{c}0, 0\\ 0, -1 \end{array}\right.\right) \text{ }\mathrm{dx},\label{Eq:C_1_step3} \\
\mathcal{C}_2 &= \frac{\rho_t}{2} \int_{0}^{\infty} x^{1/2}G_{0,2}^{2,0}\left(\left.\frac{x}{4\rho_t}\right| \frac{1}{2}, -\frac{1}{2} \right) 
\nonumber \\ & \hspace{+2.7cm}\times
G_{2,2}^{1,2}\left(\rho_t x\left|\begin{array}{c}0, 0\\ 0, -1 \end{array}\right.\right) \text{ }\mathrm{dx}\label{Eq:C_2_step3}
\end{align}
and
\begin{align}
\mathcal{C}_3 &= \frac{\rho_t}{2} \int_{0}^{\infty} x G_{0,2}^{2,0}\left(\left.\frac{x}{4\rho_t}\right| 1, -1 \right) 
\nonumber \\ & \hspace{+2.7cm}\times
G_{2,2}^{1,2}\left(\rho_t x\left|\begin{array}{c}0, 0\\ 0, -1 \end{array}\right.\right) \text{ }\mathrm{dx},
\label{Eq:C_3_step3}
\end{align}
which, by applying~\cite[ch. 2.3]{B:The_H_function}, can be analytically evaluated~as
\begin{align}
\mathcal{C}_1 &= \frac{1}{2\rho_t} G_{1,3}^{3,1}\left(\frac{1}{4\rho_t^2}\left|\begin{array}{c} -1 \\ -1, -1, 0\end{array}\right. \right),
\label{Eq:C1_s2}
\\
\mathcal{C}_2 &= \frac{1}{2\sqrt{\rho_t}} 
G_{1,3}^{3,1}\left(\frac{1}{4\rho_t^2}\left|\begin{array}{c} -\frac{1}{2} \\ -\frac{1}{2}, -\frac{1}{2}, -\frac{1}{2}\end{array}\right. \right)
\end{align} 
and
\begin{align}
\mathcal{C}_3 = \frac{1}{2\rho_t} G_{2,4}^{4,1}\left(\frac{1}{4\rho_t^2}\left|\begin{array}{c} -1, 0 \\ -1, -1, -1, 1\end{array}\right.\right). 
\label{Eq:C3_s2}
\end{align}
Finally, by substituting~\eqref{Eq:C1_s2}-\eqref{Eq:C3_s2} into~\eqref{Eq:Cs_step1}, we extract~\eqref{Eq:Cs_final}. This concludes the proof.

\balance
\bibliographystyle{IEEEtran}
\bibliography{IEEEabrv,References}
\balance

\end{document}